\documentclass[journal]{IEEEtran}
\usepackage[utf8]{inputenc}
\usepackage{xcolor}
\usepackage{subfigure}
\usepackage{color}
\usepackage{graphicx}
\usepackage{balance}
\usepackage{amsmath}
\usepackage{amsfonts}
\usepackage{amssymb}%
\usepackage{pifont}%
\usepackage{times}
\usepackage{paralist}
\usepackage{textcomp}
\usepackage[hyphens]{url}
\usepackage[font={small}]{caption}
\usepackage{units}
\usepackage{amsmath,colortbl}
\usepackage{cite,comment}
\usepackage[nolist]{acronym}
\usepackage{soul}
\usepackage{array}
\usepackage{tabularx}
\usepackage{multirow}

 \usepackage{tikz}
\usetikzlibrary{calc}

\usepackage{xcolor}
\usepackage{tabularx}
\usepackage{colortbl}
\usepackage{pgfplots}
\usepackage{mathtools}

\begin{acronym}[ACRONYM]
\acro{3GPP}{the third generation partnership project}
\acro{3D}{three dimentional}
\acro{5G}{fifth generation}
\acro{6G}{sixth generation}
\acro{AI}{artificial intelligence}
\acro{AR}{augmented reality}
\acro{AoA}{angle-of-arrival}
\acro{AoD}{angle-of-departure}
\acro{BS}{base station}
\acro{CAPEX}{capital expenses}
\acro{D-MIMO}{distributed multiple-input multiple-output}
\acro{EM}{electromagnetic}
\acro{GDOP}{geometric dilution of precision}
\acro{GNSS}{global navigation satellite system}
\acro{HMI}{human-machine interaction}
\acro{IQ}{in-phase and quadrature}
\acro{JCS}{Joint Communication and Sensing}
\acro{ISAC}{integrated sensing and communication}
\acro{JRC}{joint radar and communication}
\acro{JRC2LS}{joint radar communication, computation, localization, and sensing}
\acro{ICI}{inter-carrier interference}
\acro{IOO}{indoor open office}
\acro{IoT}{Internet of Things}
\acro{IRN}{infrastructure reference node}
\acro{ITU}{International Telecommunication Union}
\acro{KPI}{key performance indicator}
\acro{KVI}{key value indicator}
\acro{LCA}{life cycle analysis}
\acrodefplural{LCA}{life cycle analyses}
\acro{LoS}{line-of-sight}
\acro{MIMO}{multiple-input multiple-output}
\acro{mmWave}{millimeter-wave}
\acro{NLoS}{non-line-of-sight}
\acro{NR}{new radio}
\acro{NTN}{non-terrestrial network}
\acro{OFDM}{orthogonal frequency-division multiplexing}
\acro{OTFS}{orthogonal time-frequency-space}
\acro{OPEX}{operational expenses}
\acro{PRS}{positioning reference signal}
\acro{QoS}{Quality of Service}
\acro{RAN}{radio access network}
\acro{RAT}{radio access technology}
\acro{RedCap}{reduced capacity}
\acro{RF}{radio frequency}
\acro{RIS}{reconfigurable intelligent surface}
\acro{RTK}{real-time kinematic}
\acro{RTT}{round-trip-time}
\acro{RMSE}{root-mean-squared-error}
\acro{SDG}{sustainable development goal} 
\acro{SLAM}{simultaneous localization and mapping}
\acro{SNR}{signal-to-noise ratio}
\acro{SIT}{sustainability, inclusiveness, and trustworthiness}
\acro{SOTA}{state of the art}
\acro{ToA}{time-of-arrival}
\acro{TDoA}{time-difference-of-arrival}
\acro{TR}{time-reversal}
\acro{TRP}{transmission and reception point}
\acro{TXRX}[TX/RX]{transmitter/receiver}
\acro{TX}{transmitter}
\acro{RX}{receiver}
\acro{UE}{user equipment}
\acro{UN}{United Nations}
\acro{multi-RTT}{multi-cell round-trip-time}
\acro{UL-TDOA}{uplink time-difference-of-arrival}
\acro{DL-TDOA}{downlink time-difference-of-arrival}
\acro{UMi}{3D-urban micro}
\acro{UMa}{3D-urban macro}
\acro{UWB}{ultra-wide band}
\acro{FR1}{frequency range 1}
\acro{FR2}{frequency range 2}

\end{acronym}
\begin{document}

\bstctlcite{IEEEexample:BSTcontrol}
\title{6G Positioning and Sensing 
Through the Lens of Sustainability, Inclusiveness, and Trustworthiness}

\author{Henk Wymeersch, Hui Chen, Hao Guo, Musa Furkan Keskin, \\Bahare M.~Khorsandi, Mohammad H.~Moghaddam, Alejandro Ramirez, \\Kim Schindhelm, Athanasios Stavridis, Tommy Svensson, and Vijaya Yajnanarayana}

\twocolumn
\maketitle

\begin{abstract}
6G promises a paradigm shift by integrating positioning and sensing, enhancing not only the communication performance but also enabling location- and context-aware services. Historically, positioning and sensing were focused on cost and performance tradeoffs, implying an escalated demand for resources, such as radio, physical, and computational resources, for improved performance. However, 6G  expands this perspective, embracing a set of broader values, namely sustainability, inclusiveness, and trustworthiness. From a joint industrial/academic perspective, this paper aims to shed light on these important value indicators and their relationship with the conventional key performance indicators in the context of positioning and sensing.
\end{abstract}


\IEEEpeerreviewmaketitle

\section{Introduction and Motivation}%
\label{section:intro}
\Ac{ISAC} is expected to be a major differentiator of 6G when compared to previous generations \cite{liu2022integrated}. The promises of \ac{ISAC} include pervasive situational awareness by  
\emph{radar-like sensing} {(detection and tracking of objects) and \emph{non-radar-like sensing} (e.g., weather, material, and spectrum sensing)}, complemented with extremely accurate \emph{position and orientation estimation of devices}. These promises are 
{expected to materialize} thanks to a variety of technological advances, including use of the mmWave and sub-THz spectrum, \acp{RIS}, \ac{AI}, {novel} \ac{RF} hardware, etc. In turn, \ac{ISAC} will enable new applications with unprecedented demands in terms of the \acp{KPI} (e.g., accuracy, latency, coverage), such as extended reality, digital twinning, and collaborative robotics \cite{behravan2022positioning}. 

The timing of 6G is 
well-aligned with the Agenda 2030 for Sustainable Development by the \ac{UN}. Under this agenda, 17 interlinked \acp{SDG} have been defined, which serve as a ``shared blueprint for peace and prosperity for people and the planet, now and into the future'' \cite{SDG}. 
The European Hexa-X 6G Flagship project{, which includes  representatives of all stakeholders involved in the 6G value chain, ranging from vendors, operators, IT industry, and high-tech companies,}
has established that 
6G can contribute to certain SDGs \cite{hexax_d12}, through 
a holistic approach towards societal, economic, and environmental sustainability. {Hexa-X also interacted with other major 6G fora, such as the \ac{ITU}, which in its definition of 6G (called IMT-2030),  defined four overarching aspects (sustainability, ubiquitous intelligence, security/privacy/resilience, and connecting the unconnected) that act as essential design principles applicable to all usage scenarios.}

 {To extend the view of the conventional \acp{KPI} towards a more comprehensive value-driven approach,} 
 \acp{KVI} have been introduced in \cite{hexax_d12} to complement the \acp{KPI} and are able to better capture the spirit of the \acp{SDG}. 
The \acp{KVI} have been defined in three categories: \emph{sustainability}, \emph{inclusiveness}, and \emph{trustworthiness}. Hence, the 6G system should itself meet each of these \acp{KVI}, not only during the lifecycle of its components, 
but also by enabling services and applications that can, in turn, improve the \acp{KVI}. 
This \ac{KVI}-inclusive vision of 6G positioning and sensing is depicted in Fig.~\ref{fig:TheBeast}, and will be elaborated in the following sections. 

In this paper, {based on inputs from stakeholders across the value chain, including telecom vendors, large and small industry,  and academia}, 
we aim to describe and structure the \acp{KVI} for 6G  in the context of ISAC (i.e., communication, positioning, and sensing), reveal their synergies and conflicts, and propose {a methodology to} 
 quantify them  (thus effectively turning them into new \acp{KPI}). {As positioning and especially sensing has more diverse \acp{KPI} than communication in different scenarios, there are many ways to connect conventional \acp{KPI} and new \ac{KVI}-induced \acp{KPI}, providing additional opportunities (e.g., possibility to optimize KVIs) as well as challenges (e.g., more constraints to handle) to optimize 6G systems.}
 Each \ac{KVI} will also be discussed in detail, shedding light on the dual role of 6G. 
By integrating these KVIs into the architectural design of 6G ISAC, we anticipate not only novel research avenues but also the facilitation of achieving the SDGs. {While this paper focuses on the technical aspects related to the 6G KVIs, they can also be supported by policy mechanisms at international, regional, and national levels.}

\section{Performance and Value Indicators}%
\label{section:kvi}
In this section, we elaborate on the 6G use cases and the corresponding \acp{KPI}. Then, we detail the three \acp{KVI} and their relation to the \acp{KPI}.

\subsection{6G Use Cases and KPIs}

\begin{figure}
    \centering
    \includegraphics[width=1\columnwidth]{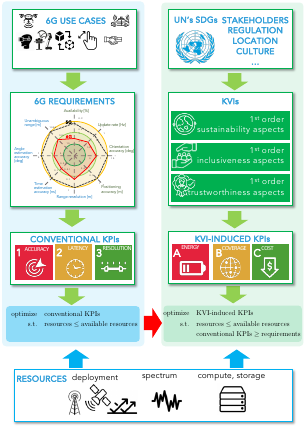}
    \caption{{A methodology for accounting for \acp{KVI} in 6G design. {The conventional design optimizes the conventional \acp{KPI} with respect to the resources, while we propose a broader approach, where first-order \acp{KVI} are transformed to \acp{KPI} and these \ac{KVI}-induced \acp{KPI} are optimized instead.}
    }}\vspace{-5mm}
    \label{fig:TheBeast}
\end{figure}

The typical 6G  use cases  can be clustered according to several verticals:  healthcare, automotive, industry, and extended reality. 
Positioning and sensing information can also be used internally by the 6G system to enhance and optimize communication functionality, for example using position information to optimize proactive resource allocation.
According to the use cases, the corresponding positioning requirements are expected to be tighter than the ones for the existing 5G standard~\cite{behravan2022positioning}, as depicted in {the left part of} Fig.~\ref{fig:TheBeast}. Moreover, new sensing requirements must be introduced in alignment with the specific use cases. 
Definition of  positioning and sensing requirements in 6G
is done through the lens of \acp{KPI},  
which are a function of the underlying radio resources and algorithms. Examples of positioning and sensing \acp{KPI} include accuracy and resolution, latency, availability, classification accuracy, {detection probability}, and update rate. 
A comprehensive exploration of the use cases and \acp{KPI} is beyond the scope of this paper, but is available in \cite{behravan2022positioning}.

\subsection{The 6G KVIs Explained}

The inception of value-based considerations in 6G, though initially introduced in \cite{hexax_d12}, has philosophical roots that can be traced to broader social awareness and responsibility~\cite{david20186g}. 
The concept of \emph{value}, as delineated by \cite{hexax_d12}, encompasses ``intangible yet essential human and societal imperatives, including growth, sustainability, trustworthiness, and inclusion.'' The operationalization of these values within the 6G framework necessitates the formulation and integration of associated criteria, in the design, functionality, and decommissioning of the system. While \cite{hexax_d12} refrained from explicit definitions of these values and the \acp{KVI}, we regard them as analogous. Our objective is to provide useful definitions for each \ac{KVI} that are comprehensive yet specific, a task that will be further expounded in the context of 6G positioning and sensing in the next section.

 \emph{Sustainability} bifurcates into environmental and economic domains.\footnote{Given that all \acp{SDG} are by definition related to sustainability, a more narrow definition is proposed.} 
 Economic sustainability pertains to practices that support long-term economic growth, balancing organizational and societal needs without undermining social, environmental, and cultural facets. Environmental sustainability was already highlighted in the 4G era \cite{fehske2011global}, 
 where life cycle analyses indicate that a holistic approach must incorporate considerations of manufacturing, operational energy consumption, recycling practices, and end-of-life treatment. However, with 6G these considerations must be considered already in the design and standardization phase.

 \emph{Inclusiveness} 
is multifaceted and aims to foster increased participation and mitigate digital divides, promoting an equitable technological landscape.
 Inclusiveness encompasses accessibility to 6G technologies, education, and facilitation in their usage, as well as assisting vulnerable demographics, such as the elderly or infants, and those marginalized due to geography, gender, culture, health, education, or other reasons.

\emph{Trustworthiness}  encompasses security (defense against deliberate attacks), 
robustness (mitigation of unintentional faults, including environmental disturbances, human errors, and system malfunctions), and privacy (unauthorized leakage of sensitive information, whether deliberate or inadvertent) \cite{fettweis20226g}. 
Notably, the anticipated pervasive utilization of \ac{AI} in 6G introduces unprecedented challenges and considerations in the realm of trustworthiness, necessitating innovative approaches.

\subsection{Relations Between KPIs and KVIs}

\begin{figure}
\centering
\includegraphics[width=0.95\linewidth]{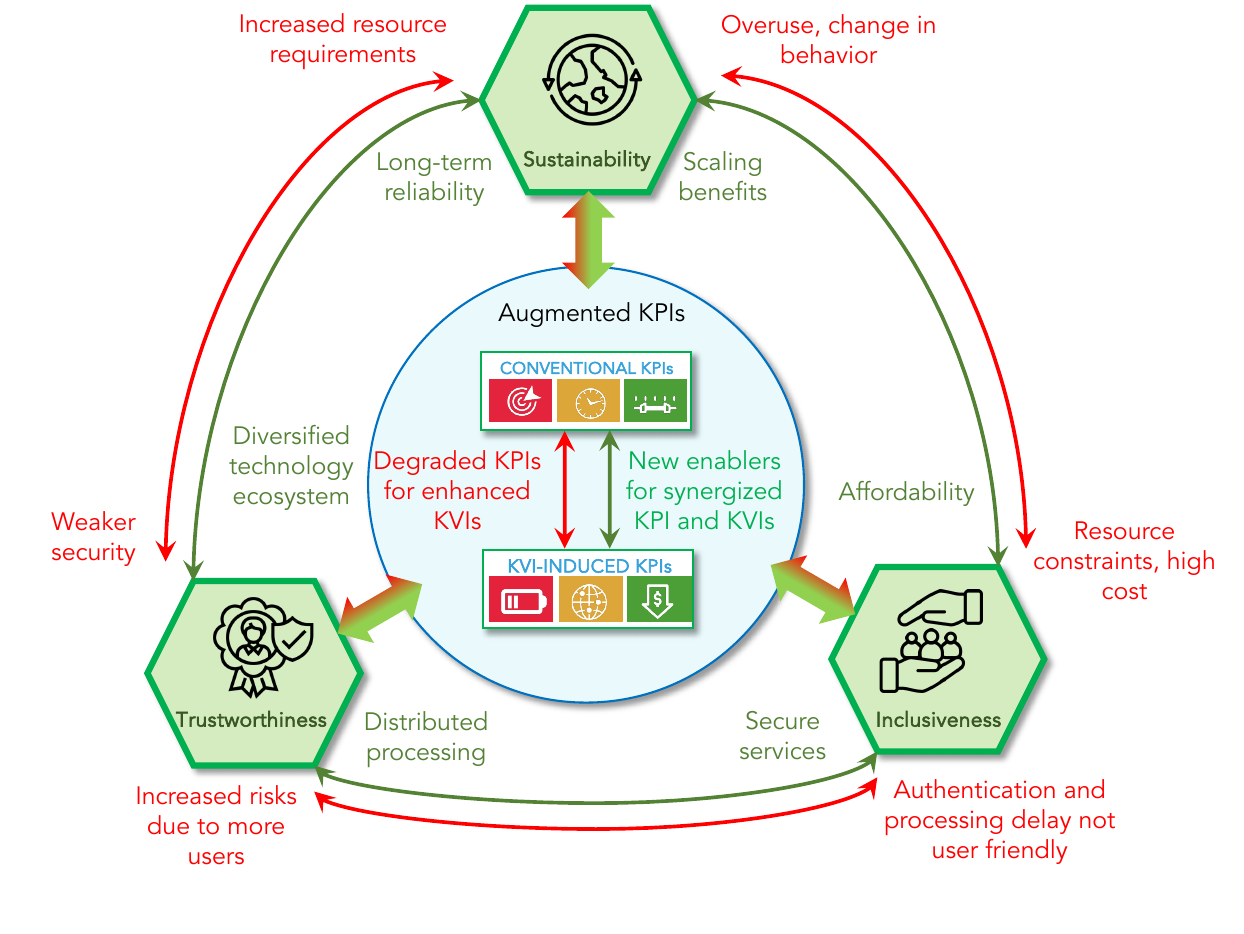}
{
\caption{Synergies (green) and trade-offs (red) among KPIs and KVIs, including higher-order effects. KPIs must be extended to quantify the KVIs in 6G design, when possible.}\vspace{-5mm}
\label{fig:relationship_kvi_kpi}}
\end{figure}

An intricate relationship exists between the conventional KPIs and  KVIs, underlined by a multifaceted interplay of trade-offs and synergies, as visually depicted in Fig.~\ref{fig:relationship_kvi_kpi}. This relationship is further elaborated below, incorporating the quantitative methodologies for KVIs and exposing the challenges emanating from potential knock-on effects.

\subsubsection{Trade-off between KPIs and KVIs}
Achieving a particular KPI might necessitate a compromise on a corresponding KVI. Pursuing heightened accuracy might demand extensive infrastructure deployment or resource consumption, undermining sustainability. Consequential impacts may manifest in reduced trustworthiness (owing to a less diversified technology ecosystem) and diminished inclusiveness (resulting from unaffordable services for specific demographics).
Conversely, elevating a KVI may cause conflicts with KPIs. The construction of a trustworthy system, albeit fostering secure services and long-term reliability, might entail additional resources or complex algorithms. This, in turn, might introduce latencies or degrade performance within the given resource constraints, affecting the associated KPIs.
 {Enhancement in one KVI may result in unintended repercussions in another KVI. For instance, to improve inclusiveness, we need to improve the sensing coverage. However, improving the sensing coverage in places where there are privacy concerns  might harm the trustworthiness (and the acceptance of the technology).}

\subsubsection{Synergy between KPIs and KVIs}
Certain scenarios reveal mutual support between KPIs and KVIs. Accurate position and map information can improve energy efficiency via so-called channel knowledge maps. 
Enhancements in positioning and sensing, coupled with broadened service reach, can promote user inclusiveness. This may, in turn,  catalyze commercialization and privacy through distributed processing, thereby enabling accurate cooperative positioning.
Trustworthiness and sustainability are valued intrinsically by users, thereby amplifying inclusiveness through wider adoption. By carefully exploiting these synergies, future networks can be designed to concurrently optimize both KPIs and KVIs, ensuring both performance objectives and broader societal benefits are achieved.
A salient instance of this synergy manifests in hardware impairment exploitation, where attributes of cost-efficient hardware (contributing to sustainability and inclusiveness) can be harnessed to enhance KPIs, such as sensing accuracy and unambiguous range \cite{PN_Exploitation_TSP_2023}.

\subsubsection{Quantification of KVIs}

While KPIs can offer quantifiable metrics for evaluating positioning and sensing performance in 6G networks, quantifying KVIs poses a formidable challenge as they often encompass essential societal values that lack a rigorous mapping to tangible metrics \cite{hexax_d12}. 
To address this challenge, a possible methodology is visualized in 
Fig.~\ref{fig:TheBeast}, where the conventional principle of optimizing \acp{KPI} subject to resource constraints, is replaced with \ac{KVI}-driven approach. Given the use case and the context (including SDGs, culture, and regulations), the first-order \ac{KVI} aspects can be identified. These first-order aspects are then mapped into so-called \emph{KVI-induced KPIs}: 
\begin{itemize}
    \item \textbf{Sustainability-induced KPIs:} Relevant KPIs include \emph{energy efficiency}, which is relatively well-defined for communication, but not for positioning and sensing, as well as \ac{CAPEX} (e.g., deployment cost) and \ac{OPEX} (e.g., power consumption of components or systems). 
    \item \textbf{Inclusiveness-induced KPIs:} Possible KPIs include \emph{coverage} that can be provided within the legacy KPI (e.g., accuracy and latency) requirements, \emph{cost} of the device or service for the end-user, \emph{accuracy} of new human-machine interfaces (e.g., via  gesture recognition).  
    \item \textbf{Trustworthiness-induced KPIs:} 
    The broad nature of trustworthiness requires metrics like \emph{position integrity} to ensure robustness against faults, and security evaluation through the \emph{probability of undetected or wrongly detected attacks and the subsequent impact}. Privacy considerations may invoke measures such as \emph{differential privacy} and \emph{mutual information} metrics.
\end{itemize}
This approach then naturally allows for optimization of the \ac{KVI}-induced \acp{KPI}, which can be optimized, still meeting the \ac{KPI} requirements, as again depicted in Fig.~\ref{fig:TheBeast}.  
While this methodology cannot fully capture the high-order effects (e.g., improved coverage may cause people to travel further and use more fossil fuels, thus reducing sustainability), it offers a first attempt for a value-oriented 6G design.

\subsection{{A Quantitative Analysis Example}}
\begin{figure}
    \centering
    \includegraphics[width=0.99\columnwidth]{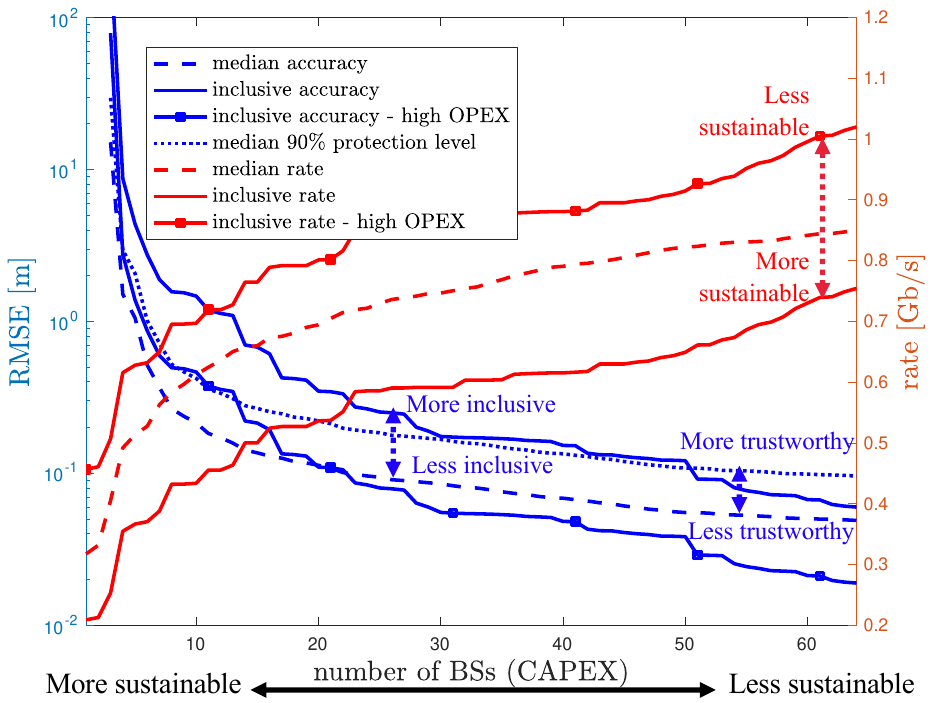}
    \caption{{
 {Quantitative analysis of KVIs with mapping to KPIs.   
    Conventional 
       \acp{KPI} (positioning accuracy and communication rate), vs.~number of \acp{BS} (representing the CAPEX). Sustainability is measured through the number of \acp{BS} and the total power consumption. Inclusiveness is measured through the 95\% (across users or locations) attainable \ac{KPI} values. Increasing the power consumption  (OPEX) $10 \times$ improves the inclusive KPIs. Trustworthiness is measured through the protection level.}
       } } \vspace{-5mm}
    \label{fig:quantitative}
\end{figure}

{Fig.~\ref{fig:quantitative} applies the methodology from Fig.~\ref{fig:TheBeast} and considers a quantitative analysis of a communication \ac{KPI} (namely rate  in Gb/s with respect to the closest \ac{BS}, computed via the Shannon–Hartley Theorem) and positioning \acp{KPI} (namely (i) a bound on the positioning \ac{RMSE} based on all \acp{BS}, computed via a Fisher information analysis, and (ii) the protection level, defined as the 90\% confidence interval in 2D, also derived from Fisher information analysis), 
as a function of the number of \acp{BS}, considering a 5.9 GHz carrier, 80 MHz bandwidth, 0.1 W transmit power per BS for pilot transmissions of 25 us per BS, and \ac{LoS} channels.}\footnote{{Source code  at \url{https://github.com/henkwymeersch/6GpositioningKVIs}}.} Users are uniformly deployed in a $4~\unit{km}^2$ area and \acp{BS} according to a regular grid within this area.
{The design should be inclusive (captured by the KPIs that can be attained for  95\% of the users, or, equivalently the deployment area),  sustainable (captured by the \ac{CAPEX}  (number of \acp{BS}) and \ac{OPEX} (the power consumption per active \ac{BS}), and trustworthy (captured by the protection level).} 
 These together comprise the \ac{KVI}-induced \acp{KPI}. 
Many insights can be drawn from this figure. 
\begin{enumerate}
    \item \textbf{KPIs have first-order impact on KVIs:} The conventional \acp{KPI} (rate and positioning accuracy), which all benefit from more BSs and more transmission power, will make the design less sustainable (CAPEX and OPEX). 
    \item \textbf{KVIs can be in trade-off:} Sustainability and inclusiveness are in contrast, so for a solution to be more sustainable it will be less inclusive and vice versa since a more inclusive design requires more resources. For instance,     
    a non-inclusive design (which serves 50\% of users{, with a target positioning RMSE}) is much less costly from a sustainability perspective than a design that is inclusive and reaches 95\% of the users {(with the same target positioning RMSE)}. 
    \item \textbf{Non-linear relations between KVIs:} 
    {Small reductions in conventional KPIs can lead to large savings in KVIs. 
    For instance, reducing the target RMSE from 0.1 m to 0.3 m, allows a $10 \times$ reduction of BS transmit power. }
    \item \textbf{Not all KPIs are equally important for KVIs:} The positioning KPIs have a much wider variation than the communication KPI, meaning that (i) positioning and sensing will likely be bottlenecks for both sustainability and inclusiveness, but also (ii) we can gain in terms of the KVIs by relaxing  selected KPIs (e.g., by allowing sensor fusion), as shown in the blue curves with a few numbers of BSs. 
\end{enumerate}

\section{The Dual Role of Positioning and Sensing from a KVI Perspective}%
\label{section:positioning}

In this section, we go deeper into each of the KVIs, to provide specific examples of how they relate to positioning and sensing in 6G. For each KVI, a dual view is taken. 
First of all, we consider how the network can operate in a way that makes positioning and sensing aligned with each value, with the help of the key 6G technologies \cite{behravan2022positioning} (including \ac{RIS}, \ac{NTN}, sidelink, \ac{AI}, \ac{D-MIMO}, and sub-THz signals). Conversely, we consider how 6G positioning and sensing, conceptualized as a service, can improve the KVIs of applications and scenarios in verticals.

\subsection{Sustainability}

Sustainability is arguably among the major concerns in 6G systems, guiding the entire lifecycle design.

\subsubsection{Sustainable Positioning and Sensing}

We consider three dimensions of sustainable design: radio resource optimization, infrastructure optimization, and the level of integration of positioning and sensing within 6G communication. 
\begin{itemize}
    \item \emph{Radio resources:} Positioning and sensing require radio resources for their operation, which consume transmit power. 
    Conservative designs based on over-provisioning should be avoided in favor of flexible and adaptive resource allocation schemes, such that energy and resource consumption can be minimized, while still (exactly) meeting the instantaneous target KPIs. Complementary to the radio resources, sleep/idle modes should be activated whenever possible to conserve energy. 

    \item \emph{Infrastructure:} Positioning and sensing generally require a more extensive infrastructure deployment than communication. 
    Such an extension is provided in 6G through two emerging technologies: \ac{D-MIMO} systems and \acp{RIS}. In \ac{D-MIMO} deployments, \acp{UE} are surrounded by a large number of energy-efficient \acp{BS}, providing not only outstanding performance in communication but also in positioning and sensing.    
   \Acp{RIS} are a class of low-energy equipment 
    that can replace/complement location anchors (e.g., \acp{BS}) and    
manipulate the wireless environment~\cite{strinati2021reconfigurable}, resulting in better propagation channels, especially in the presence of blockages. 
Similar to the radio resources, the infrastructure should be optimized, for instance, the deployment, the manufacturing, and the replacement possibility of \ac{D-MIMO} and \ac{RIS} systems, to improve sustainability under long-term target KPI requirements.
    \item \emph{Level of integration:} 
     One of the key features of 6G is to use resources and infrastructures for both positioning/sensing and communications, thereby inherently improving sustainability. 
    The integration of positioning/sensing and communications can span different levels, from sites, spectrum, and infrastructure, to waveforms and time/frequency resources. 
    While progressive integration improves sustainability, there are unavoidable trade-offs in terms of performance. Hence, stringent KPI requirements may not be suitable for the tightest possible integration. 
\end{itemize}

\subsubsection{Positioning and Sensing for Sustainability}

Positioning and sensing, through their ability to understand and digitize the physical world, provide a unique tool to enhance sustainability. 
First of all, by harnessing positioning and sensing information, data communication sustainability can be improved (e.g., context-aided communication with proactive resource allocation, beam alignment, and blockage avoidance) \cite{GMA+22}.
In addition to the more sustainable operation of communication, the ability to sense and localize has broader sustainability implications, such as earth monitoring (e.g., the ability to monitor pollution and weather){, relying on non-radar-like sensing modalities}. Recalling the verticals from Fig.~\ref{fig:TheBeast}, sustainability benefits in healthcare include the reduction in CO2 emissions thanks to remote surgery and drone deliveries{, when compared to traditional means of transporting people and goods via planes and ground vehicles}. In the automotive sector, traffic coordination and platooning {supported by 6G sensing} can be used to minimize fuel/battery consumption. In the industry vertical, 
digital twins (e.g., twins for manufacturing and autonomous supply chains, twins for sustainable food production, or twins in the context of immersive smart cities) can track the position of assets or humans {via 6G} to optimize processes, save material, and reduce waste or energy per produced item. Finally, in the realm of extended reality, the ability to collaborate virtually can lead to enormous CO2 savings, due to reduced ground and air travel.

\subsection{Inclusiveness}
In the pursuit of global digital equity, 6G should ensure
accessibility 
to all humans, irrespective of {gender, age, ability, and geographical location} \cite{ITUaccessability}. 
An integral part of this vision is to make the technology affordable, scalable, and ubiquitous. As such, positioning and sensing are the core aspects of this inclusive objective.

\subsubsection{Inclusive Positioning and Sensing}
\begin{figure}
    \centering
    \includegraphics[width=0.99\columnwidth]{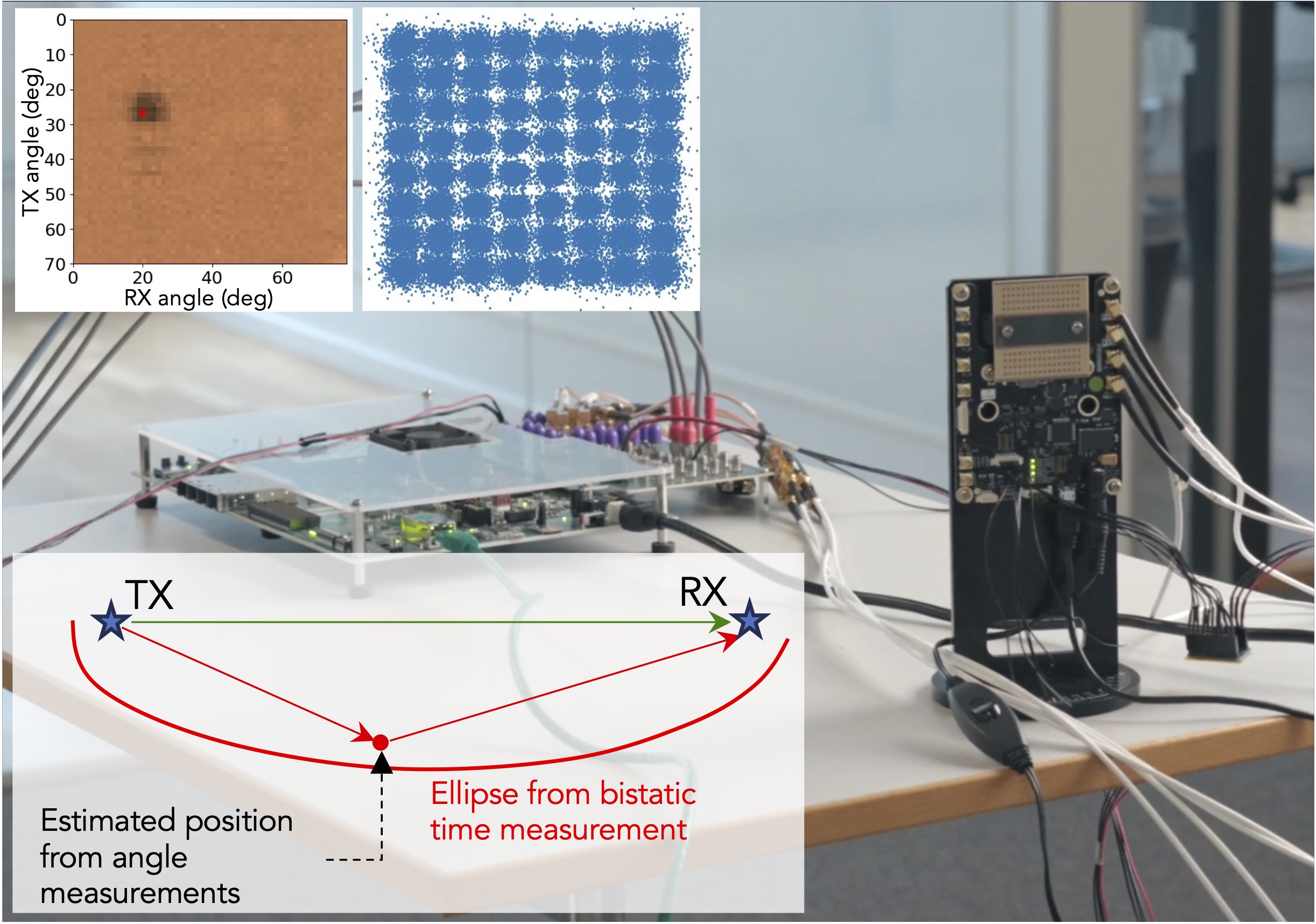}
    \caption{Proof-of-concept for joint communication and sensing, showing how existing communication infrastructure and signals can be repurposed for sensing, in support of  sustainability and inclusiveness.      
    The hardware comprises Sivers semiconductors EVK06002 as \ac{TX} and \ac{RX}, each with 
     1×16 arrays. Standard 5G waveform with 120 kHz subcarrier spacing, 800 MHz bandwidth, 69 GHz carrier frequency, and 64 QAM modulation are employed, {yielding a  maximum  rate of 560 Mb/s}.  
    Besides the data transmission (top middle),  beam sweeping (top left) provides bearing measurement of the passive target. Bistatic time measurements provide a sensing ellipse to further improve the target position estimate (bottom left). {The sensing resolution is 40 cm and the sensing duration is 6.25 ms (2800 beam combinations and 2 OFDM symbols per beam pair) \cite{hexax_d33}.}
    }
    \label{fig:inclusiveness}
\end{figure}

Positioning and sensing, embedded in the network architecture, can be facilitated by network deployment across all geographical terrains. This is feasible through a combination of several developments: 
the reuse of communication resources and infrastructure for multi-purpose functionality, ubiquitous connectivity, and cooperative networks.
\begin{itemize}
    \item \emph{Multi-purpose functionality:} 
    The infrastructure for providing communication and network services will be repurposed 
    for positioning and sensing functions. 
    A proof-of-concept for this dual-purpose application is illustrated in Fig.~\ref{fig:inclusiveness}, where communication signals are used to track a person.  
    \item \emph{Ubiquitous connectivity}: 
    The incorporation of \acp{NTN} will significantly extend the coverage of 6G networks to remote or difficult-to-reach areas, ensuring that geographical barriers do not limit access to vital communication or sensing services. 
    Similarly, \acp{RIS} also enhance and enable accurate and efficient positioning and sensing in various scenarios, largely extending the coverage of services~\cite{chen2023riss}.
    Consequently, ubiquitous connectivity-enabled positioning is poised to significantly augment the inclusiveness of the 6G network by enabling uninterrupted connectivity regardless of the users' proximity to the traditional network infrastructure.
    \item \emph{Cooperative networks}:
    Sidelink supports direct communication between devices, bypassing the centralized network infrastructure. This capability can facilitate the creation of localized communication networks, extending connectivity and service availability in scenarios where conventional network coverage may be absent or limited, such as in rural, remote, or disaster-struck areas. {Such a cooperative approach makes positioning and sensing tasks to be completed in a distributed manner, largely extending the coverage and reducing the cost of the provided services.}
\end{itemize}
These three aspects underscore how 6G technology will be instrumental in breaking down existing barriers in network access and functionality, demonstrating a firm commitment to creating a truly inclusive, global digital ecosystem.

\subsubsection{Positioning and Sensing  for Inclusiveness}
Inclusiveness in 6G networks is not only a macro-level objective but also addresses the accessibility challenges encountered at the micro-level of individual human-machine interactions. Positioning and sensing can play a crucial role in this context. On the one hand, advancements in sensing technology will enable systems that can interpret and respond to gestures, which benefits individuals who face challenges in traditional interaction modalities. Such a transformation can redefine the nature of human-machine interaction, making it more inclusive and accessible. 
On the other hand, intelligent monitoring, especially in critical societal domains such as elderly care, patient supervision, and infant care, emerges as a domain where sensing can be a game-changer{, by advanced gesture or posture classification or breathing monitoring based on micro-Doppler}. Such integrative applications promise to redefine caregiving, providing options characterized by precision, real-time feedback, and remote monitoring. These developments serve to enhance the quality of life for these demographic segments, underlining 6G's commitment to be genuinely inclusive and beneficial to all of society.
Referring back to the proof-of-concept demonstration from Fig.~\ref{fig:inclusiveness}, a person can be tracked in a cluttered environment with the aid of {6G} communication signals and infrastructure, negating the need for additional equipment or invasive monitoring.

\subsection{Trustworthiness} \label{sec:Trustworthiness}
Ensuring the robustness, security, and privacy of 6G positioning and sensing must be a priority in the design of the overall 6G system, given the safety-critical nature of the verticals. 
This section outlines challenges and approaches related to the trustworthiness of positioning and sensing in 6G.

\subsubsection{Trustworthy Positioning and Sensing}
We deconstruct trustworthiness into its constituent elements, such as robustness, security, and privacy, before discussing the influence of, e.g., \ac{AI}, on them separately. 
\begin{itemize}
    \item 
\emph{Robustness:} Robust positioning and sensing are primarily based on diversity, relying on a large set of measurements from independent technologies, observations, or dimensions, to provide redundancy for detecting and eliminating faults. This approach is common in \ac{GNSS}, where for instance, aviation applications demand protection levels with a high degree of certainty, even in the presence of faults. The 6G system itself can provide inherent redundancy, via diverse measurements (e.g., not only \ac{TDoA}, but also \ac{AoA}, \ac{AoD}, carrier phase, and perhaps Doppler), diverse location references (e.g., using many access points in \ac{D-MIMO}), and multi-sensor fusion (e.g., relying on a combination of 6G sensing with vision). When combined with integrity monitoring, 6G can provide high performance with guaranteed robustness \cite{Reid_2019}.

\item \emph{Security:} Vulnerabilities exist in classical positioning technologies (e.g., \ac{GNSS} and \ac{UWB}), where attackers can perform jamming (blinding the receiver, leading to service interruption), meaconing (retransmission of legitimate signals), or spoofing (transmission of false signals)\cite{goztepe2021localization}.
Spoofing can be mitigated by cryptographic countermeasures, while jamming can be mitigated by directional nulling at the receiver. Attacks on radar sensing include jamming, altering electromagnetic properties, deception, masking, and imitation. Adaptive waveform design and frequency hopping help correct target range or velocity errors.
Extrapolating these concepts to 6G, it is clear that each measurement type (delay, angle, Doppler), each piece of hardware (\ac{BS}, \ac{RIS}, \ac{UE}), and each waveform have potential security weaknesses that can compromise positioning and sensing. 
An example of a  positioning and sensing attack in a 6G context is shown in Fig.~\ref{fig:attacks}, where an attacker manipulates the \ac{TX} beamforming, which leads to perceived high-power paths at the \ac{RX} with modified \ac{AoD} (with limited knowledge at the attacker) or \ac{AoA} (with complete knowledge at the attacker).

\item \emph{Privacy:}
 Privacy protection in the area of location tracking of humans is already crucial for 5G and comes even more into focus with 6G's higher positioning accuracy (including its opportunities for cross-platform fusion of tracking information, and exposure framework for internal and external use). Position information that can be easily used for behavioral profiling must be secured from unauthorized access on all levels (including physical-layer security). Moreover, not only tracking of humans is possible but tracking of objects and assets as well. In corporate environments, where asset tracking is used to monitor and optimize processes, this process information becomes worthwhile protecting as well. Technological protection includes solutions such as active cloaking, reminiscent of techniques in electronic warfare.  
\end{itemize}
{In the context of the trustworthiness of 6G, the advent of AI has the potential to instigate novel attacks, exploiting latent system vulnerabilities. Conversely, AI can fortify system security and privacy by innovating newly learned protocols or waveforms. { However, the opaque nature of AI mechanisms demands rigorous and transparent scrutiny, especially for safety and mission-critical tasks. Formal verification methods such as model checking which are commonly employed in hardware system design can be used to verify complex AI models. Some of the properties that these models verifies could provide necessary explanations for these tasks.}

\begin{figure}
    \centering
    \includegraphics[width=0.99\columnwidth]{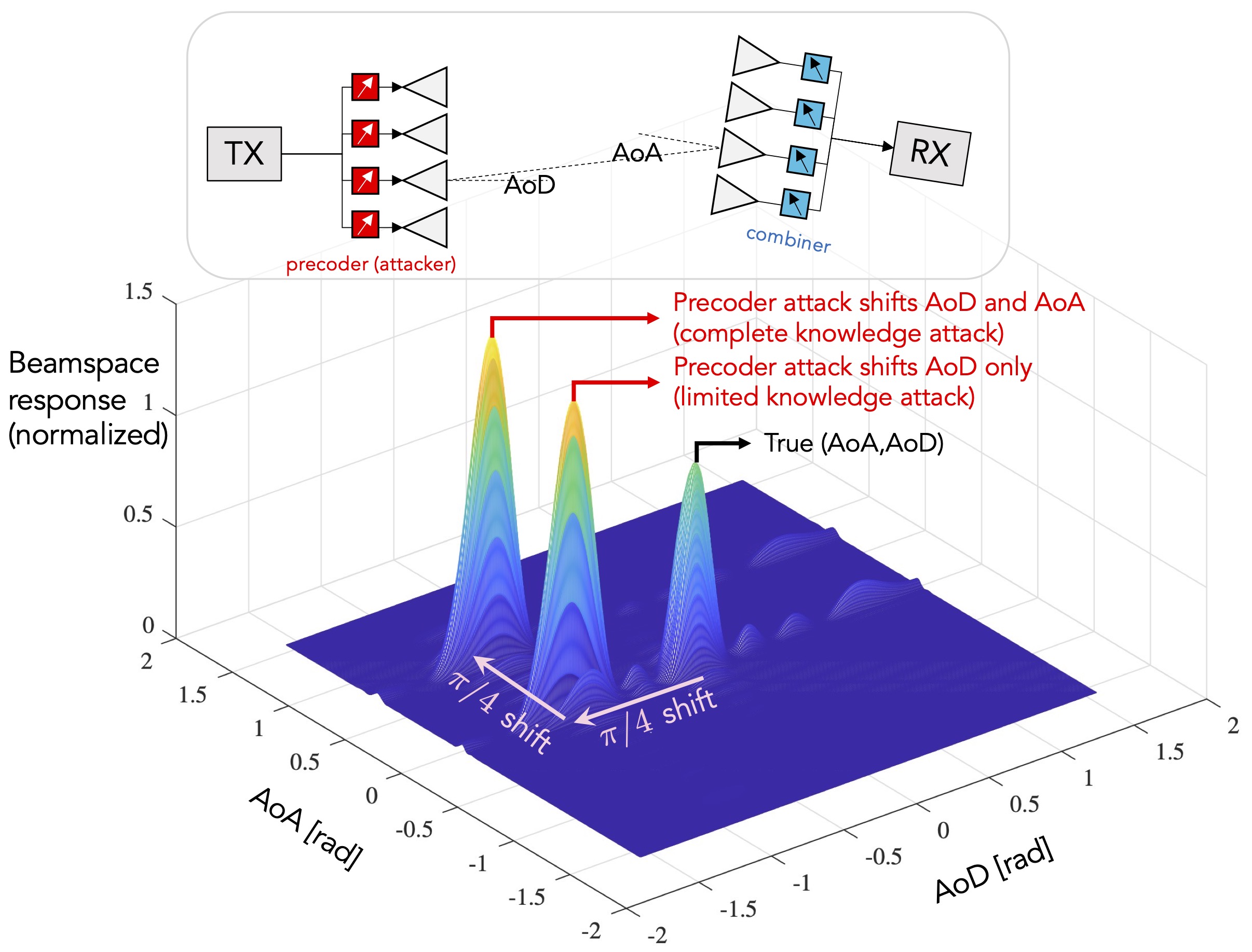}
    \caption{A 6G ISAC attack example, where a transmitter modifies its beamforming vector to fool an analog/hybrid receiver, {each with 16 elements} into believing there are additional (strong) paths at {controlled} \ac{AoD} or \ac{AoA}, shifted $\pi/4$ in each domain with respect to the \ac{LoS} path at \ac{AoD} and \ac{AoA} of $0$ radians. }
    \label{fig:attacks}\vspace{-5mm}
\end{figure}

\subsubsection{Positioning and Sensing for Trustworthiness}
The ability to localize users and objects with a high degree of accuracy can support applications that rely on trustworthiness. First of all, in terms of robustness, 6G network/system will act as an additional sensor, complementing and verifying existing sensors (e.g., camera, GPS, lidar, radar, inertial measurement unit). This will benefit all safety-critical services (including the corresponding communication), where incorrect location information may lead to harm. Secondly, security functions can be based on accurate location information or biometric 6G sensing data can be employed in access control or payment services. 
Given the built-in encryption and security frameworks, 6G data is poised to receive greater trust than other sensory inputs, driving the emergence of novel applications. Lastly, surveillance and crowd control applications are envisioned to benefit immensely from the sensory data facilitated by 6G.

\newcolumntype{P}[1]{>{\centering\arraybackslash}m{#1}}
\newcolumntype{Q}[1]{>{\arraybackslash}m{#1}}
\begin{table*}[]
    \centering
    {
    \resizebox{2\columnwidth}{!} {
    \begin{tabular}{|Q{1.1cm}|Q{2cm}|Q{2cm}|Q{2cm}|Q{2cm}|Q{2cm}|Q{2cm}|}
    \hline    
     {\cellcolor{blue!10}\textbf{Enablers$\downarrow$}} 
     & {\cellcolor{green!10}\textbf{Pro \newline Sustainability}}  &   {\cellcolor{red!10}\textbf{Con \newline Sustainability}} & {\cellcolor{green!10}\textbf{Pro \newline Inclusiveness}} & {\cellcolor{red!10}\textbf{Con \newline Inclusiveness}} 
     & {\cellcolor{green!10}\textbf{Pro \newline Trustworthiness}} &  {\cellcolor{red!10}\textbf{Con \newline Trustworthiness}}\\ 
    \hline
    RIS & low energy consumption for P\&S compared to BS & additional RIS deployment costs &  low cost compared to additional BSs and extended coverage  & focused radio wave exposure concerns; limited coverage extension  compared to BSs  & RIS links provide redundancy, increasing P\&S robustness and integrity & The RIS itself attacks the network, leading to spoofed P\&S measurements  \\
    \hline
    AI & can learn to optimize CAPEX and OPEX for P\&S & training and inference complexity leads to high energy consumption &  can support P\&S functions to serve non-connected users, e.g., remote monitoring & expensive infrastructure could limit accessibility to certain regions or groups  & AI can be used to detect anomalies in P\&S measurements or received signals & AI decisions are not always explainable; AI can be used to learn new attacks \\
    \hline
    NTN & 1 satellite or high altitude platform station can replace several base stations in P\&S &  additional deployment costs and difficult to decommission & extreme P\&S coverage enhancements in under-served regions
    &  possibly high service fees; scarce spectrum limits sensing resolution
    & very difficult to tamper with NTN P\&S signals
    &  NTN signals are easier to jam due to their lower power \\
    \hline
    D-MIMO & short distance to user leads to reduced power consumption and more geometric diversity & high density deployment and additional cabling required & uniform P\&S coverage and flexible architecture & mainly for dense areas with many users and good fiber network & high level of redundancy due to many possible links, increasing robustness and integrity of P\&S& possibility for pervasive and continuous tracking of users leads to privacy concerns. \\
    \hline
    Sub-THz & extreme focusing provides high SNR links; can replace cabled solutions & low efficiency of power amplifiers and limited use cases
    & new sensing services (e.g., imaging) at low cost with needed dedicated equipment & mainly for nearly static links & directional links avoid eavesdropping and limit impact of jamming
    & limited robustness due to links drop under mobility or environmental effects \\
    \hline
    Sidelink &  provides connectivity for sensing with fewer BSs & extensive coordination signaling needed; draining user device batteries more quickly & can provide P\&S services with limited or no infrastructure deployment & limited to relative P\&S and sensing without infrastructure & local communication and coordination of P\&S does not leak into the network & lack of central unit limits security mechanisms \\ 
    \hline
    \end{tabular}
                    }}
     \vspace{1mm}
    \caption{{Selected enablers for 6G positioning and sensing (P\&S) and example benefits (Pro) and drawbacks (Con) in terms of the impact on the \acp{KVI}, based on analysis in the Hexa-X project \cite{hexax_d33}.}}

    \label{tab:prosandcons}
    \vspace{-3mm}
\end{table*}

\subsection{Impact of 6G Enablers on the KVIs}
{To conclude this section, we offer an overview of several key technological enablers for 6G positioning and sensing, \cite{behravan2022positioning} (namely, \ac{RIS}, \ac{NTN}, sidelink, \ac{AI}, \ac{D-MIMO}, and sub-THz signals) and list \emph{exemplifying benefits and drawbacks} in terms of the \acp{KVI} (see Table~\ref{tab:prosandcons}), based on analyses conducted in Hexa-X \cite{hexax_d33}.
The precise trade-offs and synergies and associated benefits and drawbacks depend on the specific use case and its requirement, so the entries in Table.~\ref{tab:prosandcons} should be interpreted as examples, not guidelines or general conclusions. 
It is also important to note that several of the benefits and drawbacks emanate from higher-order effects, underscoring the intricate and multifaceted nature of 6G system design. }

\section{Outlook}%
\label{section:outlook}

The evolution of precise positioning and sensing for 6G \ac{ISAC} presents a set of challenges and opportunities. As this paper has underscored from both academic and industrial perspectives, the next generation of digital communication is not merely about advancing the conventional \acp{KPI}, but also to forge a digital ecosystem that is sustainable, inclusive, and trustworthy, in line with the UN's \acp{SDG}. 
We have shown that these values should be related to \acp{KVI}, which in turn can be quantitatively mapped to new \acp{KPI}. Both synergies and trade-offs will occur, and higher-order effects should be considered. 
For each of the \acp{KVI}, this paper has revealed the intricate nature of 6G positioning and sensing, both to make positioning and sensing coalesce with the \acp{KVI}, and to provide services that enhance the \acp{KVI}. 

As we stand on the cusp of the 6G era, it has become clear that the adoption of a holistic approach is imperative. As researchers, engineers, scientists, and stakeholders, our task is not only to innovate, but also to ensure that the digital future is sustainable, inclusive, and trustworthy.

\section*{Acknowledgments}
This work was supported, in part, by the European Commission through the H2020 project Hexa-X (Grant Agreement no. 101015956), and by Swedish Research Council (VR) 2023-00272. The authors are grateful to Hamed Farhadi (Ericsson) for his feedback.

\balance 
\bibliographystyle{IEEEtran}
\bibliography{IEEEabrv,reference}

\vspace{-10mm}

\begin{IEEEbiographynophoto}
{Henk Wymeersch} is a Professor with the Department of Electrical Engineering at Chalmers University of Technology, Sweden, in the area of radio localization and sensing. 
  \end{IEEEbiographynophoto}
   \vspace{-1.2cm}
   \begin{IEEEbiographynophoto}
{Hui Chen} is a postdoc with the Department of Electrical Engineering at Chalmers University of Technology, Sweden, focusing on mmWave/THz localization and sensing.
  \end{IEEEbiographynophoto}
 \vspace{-1.2cm}
   \begin{IEEEbiographynophoto}
{Hao Guo} is a Postdoc with the Department of Electrical Engineering at Chalmers University of Technology, Sweden, and also a Postdoctoral Visiting Scholar with Electrical and Computer Engineering Department, New York University Tandon School of Engineering, Brooklyn, NY, USA. His research is on integrated sensing and communications.
  \end{IEEEbiographynophoto}

  \vspace{-1.2cm}
   \begin{IEEEbiographynophoto}
{Musa Furkan Keskin} is a Research Specialist at Chalmers University of Technology, Sweden, focusing on integrated sensing and communications with hardware impairments in beyond 5G/6G systems.
  \end{IEEEbiographynophoto}

  \vspace{-1.2cm}
   \begin{IEEEbiographynophoto}
{Bahare M.~Khorsandi,} is a Core Network Specialist at Nokia Strategy and Technology in Munich, Germany, working on 6G network architecture design.
  \end{IEEEbiographynophoto}

  \vspace{-1.2cm}
   \begin{IEEEbiographynophoto}
{Mohammad H.~Moghaddam} is a Research Specialist and joint communication and sensing expert at QAMCOM Research and Technology AB, Gothenburg, Sweden.  
  \end{IEEEbiographynophoto}

  \vspace{-1.2cm}
   \begin{IEEEbiographynophoto}
{Alejandro Ramirez} is a Senior Key Expert at Siemens' corporate research unit in Munich, Germany, working on wireless communication and localization.
  \end{IEEEbiographynophoto}

  \vspace{-1.2cm}
   \begin{IEEEbiographynophoto}
{Kim Schindhelm}  is a Research Scientist at Siemens' corporate research unit in Munich, Germany, working on localization. 
  \end{IEEEbiographynophoto}

  \vspace{-1.2cm}
   \begin{IEEEbiographynophoto}
{Athanasios Stavridis} is a Senior Researcher at Ericsson Research, Ericsson, Sweden, working in the field of signal processing for wireless communication and sensing. 
  \end{IEEEbiographynophoto}
  
  \vspace{-1.2cm}
   \begin{IEEEbiographynophoto}
{Tommy Svensson} is a Professor with the Department of Electrical Engineering at Chalmers University of Technology, Sweden, in the area of wireless systems.
  \end{IEEEbiographynophoto}
  \vspace{-1.2cm}
   \begin{IEEEbiographynophoto}
{Vijaya Yajnanarayana} is a Master Researcher at Ericsson Research, India, working in the area of radio signal processing and artificial intelligence.
  \end{IEEEbiographynophoto}

\end{document}